\documentclass{PoS}
\usepackage{amsmath}

\title{Progress towards precision measurements of $\beta$-decay correlation 
  parameters using atom and ion traps}

\ShortTitle{Progress towards $\beta$-decay correlation measurements using atom and ion traps}

\author{\speaker{D. Melconian},$^{ab}$ R.S. Behling,$^{ac}$ B. Fenker,$^{ab}$ 
  M. Mehlman,$^{ab}$ P.D. Shidling,$^a$, 
  M. Anholm,$^{de}$, D. Ashery,$^f$ J.A. Behr,$^d$, A. Gorelov,$^d$, 
  G. Gwinner,$^g$  K. Olchankski,$^d$ and S. Smale$^d$\\
\llap{$^a$}Cyclotron Institute, Texas A\&M University\\
College Station, TX 77843-3366 USA\\
\llap{$^b$}Department of Physics \& Astronomy, Texas A\&M University\\
College Station, TX 77843-4242 USA\\
\llap{$^c$}Department of Chemistry, Texas A\&M University\\
College Station, TX 77843-3012 USA\\
\llap{$^d$}TRIUMF\\
Vancouver, BC V6T 2A3 Canada\\
\llap{$^e$}Department of Physics \& Astronomy, University of British Columbia\\
Vancouver, BC V6T 1Z1 Canada\\
\llap{$^f$}School of Physics and Astronomy, Tel Aviv University\\
Tel Aviv, Israel\\
\llap{$^g$}Department of Physics \& Astronomy, University of Manitoba\\
Winnipeg, MB R3T 2N2 Canada\\
        E-mail: \email{dmelconian@physics.tamu.edu}}

\abstract{The correlations of the decay products following the $\beta$ 
  decay of nuclei have a long history of providing a low-energy probe of 
  the fundamental symmetries of our universe.  Over half a century ago, 
  the correlation of the electrons following the decay of polarized 
  \tsups{60}Co demonstrated that parity is \emph{not} conserved in weak 
  interactions.  Today, the same basic idea continues to be applied to 
  search for physics beyond the standard model: make precision measurements 
  of correlation parameters and look for deviations compared to their 
  standard model predictions.  Efforts to measure these parameters to the
  0.1\% level utilizing atom and ion trapping techniques are described.}

\FullConference{X Latin American Symposium on Nuclear Physics and 
  Applications (X LASNPA),\\
  1-6 December 2013\\
  Montevideo, Uruguay}


\newcommand{\tsups}[1]{\textsuperscript{#1}}

\newcommand{\trinat}{{\scshape Trinat}}
\newcommand{\triumf}{{\scshape Triumf}}
\newcommand{\tamutrap}{{\scshape Tamutrap}}

\begin{document}

\section{Introduction}
The primary goal of the Texas A\&M University trap (\tamutrap) and 
\triumf{} Neutral Atom Trap (\trinat) facilities is to search 
for new physics by pushing the precision frontier.  Precision measurements 
of $\beta$-decay correlations of certain nuclei are sensitive to physics 
beyond the standard model. In particular, pure Fermi decays (which are 
predicted to be mediated by a vector interaction) are sensitive to possible 
admixtures of a weak scalar current via the $\beta-\nu$ correlation parameter 
$a_{\beta\nu}$.  Measuring this correlation via the shape of the $\beta$-delayed 
proton energy spectrum from the decay of isospin $T=2$ nuclei will be the 
focus of initial program for the \tamutrap{} facility.  The isobaric analogue 
decay of mixed transitions, such as \tsups{37}K, are also sensitive to a 
variety of new physics (e.g.\ right-handed currents, tensor interactions, 
second-class currents, \ldots) just like the neutron.  The \trinat{} 
collaboration has a mature program performing precision $\beta$-decay 
experiments using a magneto-optical 
trap~\cite{trinczekPRL,gorelovPRL,melconianPLB,pitcairn2009}.  We have 
developed the 
ability to trap and highly-polarize \tsups{37}K using atomic techniques, and 
have recently taken data for the first-ever measurement of the 
$\beta$-asymmetry parameter, $A_\beta$, from 
laser-cooled atoms.  The goal for all of these correlation measurements, both 
by \tamutrap{} and \trinat, is to be complementary to high-energy searches 
which may be realized when reaching a precision of 
0.1\%~\cite{profumo,cirigliano}.


\section{Ion Trapping with \tamutrap}
We are in the process of building and commissioning a unique Penning trap 
system at the Cyclotron Institute on the campus of Texas A\&M University.  Our 
initial program will precisely measure the shape of the $\beta$-delayed 
proton energy spectrum from the decay of isospin $T=2$ nuclei.  Being a 
pure Fermi transition between $0^+$ states, the angular distribution of these 
decays is given simply by~\cite{JTWa,JTWb}
\begin{align}
  dW &= dW_0\,\xi\Big(1+
  a_{\beta\nu}\frac{\vec{p}_e\cdot\vec{p}_\nu}{E_eE_\nu}+
  b\frac{m_e}{E_e}\Big)\label{eq:JTW-Fermi}
  \intertext{with}
  dW_0&=\frac{G_F^2|V_{ud}|^2}{(2\pi)^5}p_eE_e(A_0-E_e)^2\,F(Z,E_e)
  \,dE_ed\Omega_ed\Omega_\nu,\label{eq:beta-decay}
\end{align}
where $E_e$ and $p_e$ are the energy and momentum of the $\beta$, $A_0$ is the 
maximum energy available to the leptons and $F(Z,E_e)$ is the Fermi function.  
The Fermi coupling constant is $G_F/(\hbar c)^2=1.16639(1)
\times10^{-5}~\mathrm{GeV}^{-2}$ (taken from $\mu$ decay) and $V_{ud}$ is the 
up-down element of the Cabibbo-Kobayashi-Maskawa (CKM) mass-mixing matrix.  
The correlation parameters $a_{\beta\nu}$ and $b$ depend on the form of the 
weak interaction and are particularly sensitive to scalar currents.  

One of the most precise limits to date on possible scalar contributions to the 
predominantly $V-A$ form of the weak interaction was made by observing the 
energy spectrum of the $\beta$-delayed proton following the superallowed 
decay of \tsups{32}Ar~\cite{adelberger,garcia32Ar}.  This 0.5\% measurement 
of the combination of $\beta-\nu$ and Fierz correlation parameters, 
$\tilde{a}_{\beta\nu}\equiv a_{\beta\nu}/(1+b\langle E_e\rangle)$, was made by 
observing only the proton energy in singles, ignoring any information from 
the $\beta$. Using the favourable source conditions of a Penning trap, we plan 
to improve on this type of measurement by observing the $\beta$ in coincidence 
with the proton.  Such a measurement will have an enhanced sensitivity as well 
as being less susceptible to backgrounds.  In order to obtain near $4\pi$ 
collection of the protons, we have designed our cylindrical Penning trap to 
be the world's largest, having an unprecedented inner trap-electrode diameter 
of $180~\mathrm{mm}$.  Additional scientific capabilities planned for this 
system are $ft$-values, lifetime measurements, mass measurements, and 
providing an extremely pure, low-energy radioactive ion beam for various 
other applications.

\begin{figure}\centering
  \includegraphics[width=0.985\textwidth]{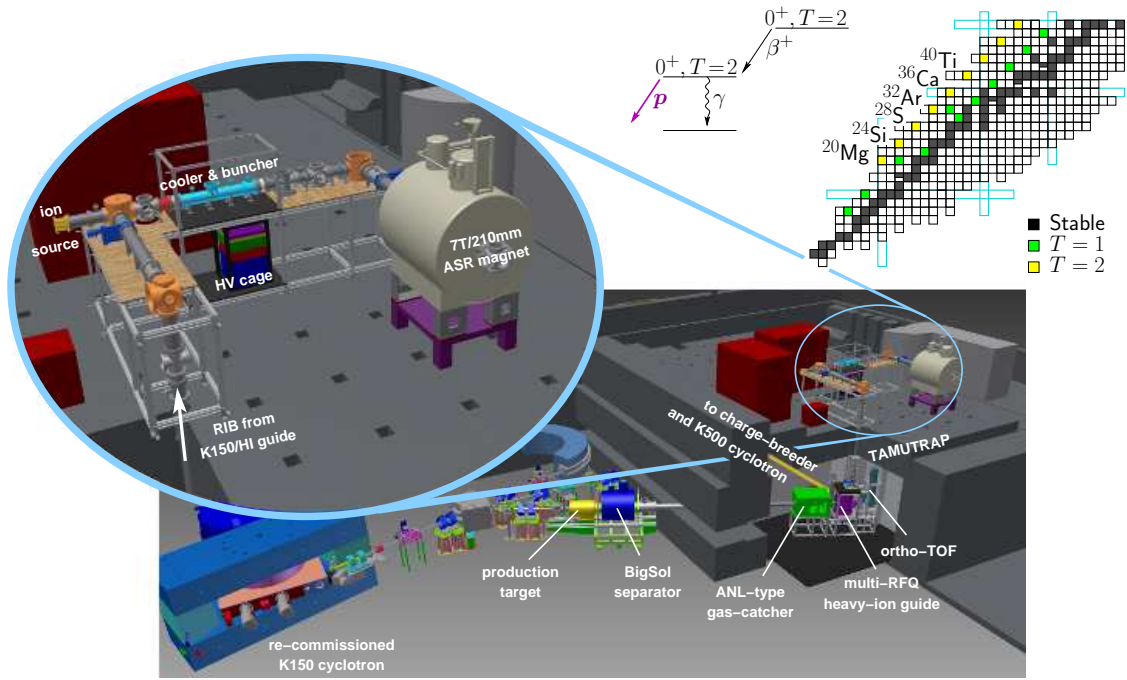}
  \caption{Layout of the \tamutrap{} facility at the Cyclotron Institute, 
    Texas A\&M University (see text).  The initial program for this Penning 
    trap system will measure the $\beta$-delayed protons from the decay of 
    a number of proton-rich $T=2$ nuclei as shown by the chart of nuclides 
    (top-right).\label{fig:tamutrap-layout}}
\end{figure}

The general layout of the \tamutrap{} facility and how it fits within the 
T-REX upgrade of the Cyclotron Institute are shown in 
Fig.~\ref{fig:tamutrap-layout}.  Radioactive 
ions will be produced via fusion evapouration reactions with a \tsups{3}He gas 
target using the high-intensity primary beams available from the K150 cyclotron. 
The reaction products will be separated following the production target using 
BigSol, a large-acceptance, 7~T solenoidal magnet.  The secondary beam will 
be stopped in the gas-catcher of the heavy-ion guide system, which can then 
send the rare ions at 15~keV beam energy either to a charge-breeder and the 
K500 for re-acceleration, or up to the \tamutrap{} facility. 
The main components of the \tamutrap{} facility are: a 
radio-frequency quadrupole (RFQ) Paul trap used to cool and bunch the ions, 
a purification Penning trap for isobar separation, and 
a novel measurement Penning trap for precision $\beta$-decay 
measurements~\cite{mehlman-NIMA}.  

Currently, \tamutrap{} is in an ongoing state of design, fabrication, and 
assembly.  Production calculations for the species of interest have been 
performed using LISE and the predicted beam capabilities of the K150 
cyclotron, indicating what rates might ultimately be expected (see 
Table~\ref{table:rates}).  To complement this, initial production experiments 
have been performed using the K500 Cyclotron in combination with the MARS 
spectrometer, in order to verify the cross-sections calculated in LISE.  The 
data for this work is still being examined.  Efficiency and rate-handling 
estimations for the entire beamline following the target chamber have been 
considered, yielding a total predicted efficiency of around 0.8\%.  Other 
simulation work that is currently ongoing includes geometric efficiency 
calculations for the proposed measurement and initial simulations and trial
 analysis of the decay of interest.

\begin{table}\centering
  \begin{tabular}{rclrc@{\ }clrr@{.}lrr@{}l}
    \hline\hline
    \multicolumn{3}{c}{}            & \multicolumn{4}{c}{}        & \multicolumn{3}{c}{Calculated}          & \multicolumn{3}{c}{Estimated}          \\[-0.85em]
    \multicolumn{3}{c}{Radioactive} & \multicolumn{4}{c}{Primary} & \multicolumn{3}{c}{}                    & \multicolumn{3}{c}{}                   \\[-0.85em]
    \multicolumn{3}{c}{}            & \multicolumn{4}{c}{}        & \multicolumn{3}{c}{cross-section}       & \multicolumn{3}{c}{production rate}    \\[-0.85em]
    \multicolumn{3}{c}{ion beam}    & \multicolumn{4}{c}{beam}    & \multicolumn{3}{c}{ }                   & \multicolumn{3}{c}{}                   \\[-0.85em]
    \multicolumn{3}{c}{}            & \multicolumn{4}{c}{}        & \multicolumn{3}{c}{[$\times10^{-3}$ mb]} & \multicolumn{3}{c}{[$\times10^5$ pps]} \\
    \hline
    &\tsups{20}Mg &&& \tsups{20}Ne & @ 24~MeV/u &&&  $16$&$2$  && $14$&    \\
    &\tsups{24}Si &&& \tsups{24}Mg & @ 23~MeV/u &&&  $15$&$5$  &&  $6$&$.5$\\
    &\tsups{28}S  &&& \tsups{28}Si & @ 23~MeV/u &&&   $4$&$5$  &&  $1$&$.5$\\
    &\tsups{32}Ar &&& \tsups{32}S  & @ 21~MeV/u &&&   $7$&$3$  &&  $1$&$.4$\\
    &\tsups{36}Ca &&& \tsups{36}Ar & @ 23~MeV/u &&&   $6$&$3$  &&  $2$&$.5$\\
    &\tsups{40}Ti &&& \tsups{40}Ca & @ 23~MeV/u &&&   $1$&$7$  &&  $0$&$.7$\\
    \hline\hline
  \end{tabular}
  \caption{Calculated production rates of the $T=2$ nuclei of interest for 
    the initial program of \tamutrap.  All isotopes are planned to be 
    produced using fusion-evapouration reactions with a \tsups{3}He gas 
    target.\label{table:rates}}
\end{table}

\begin{figure}\centering
  \includegraphics[width=0.7\textwidth]{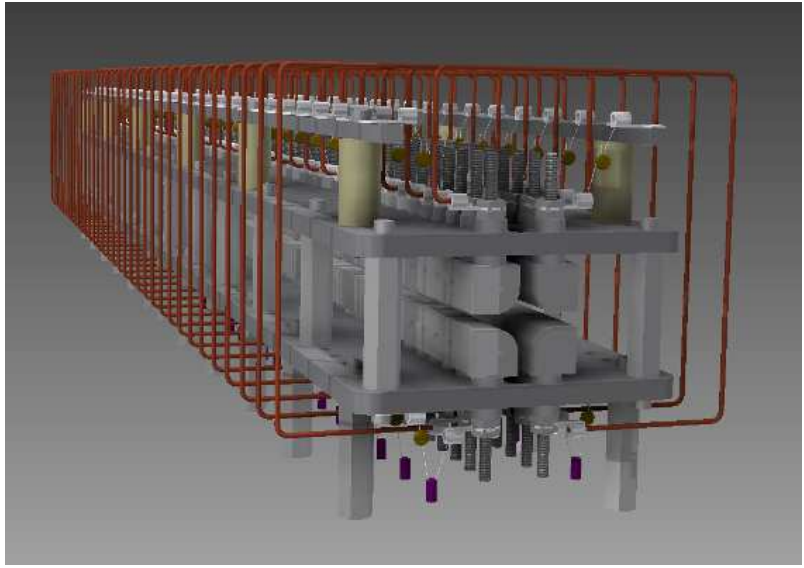}
  \caption{AutoCAD design of the RFQ cooler and buncher.  The 15~keV ions from 
    the heavy ion guide will be cooled via collisions with the 
    $\sim10^2~\mathrm{atm}$ He buffer gas, and collect at the end of the 
    28-electrode structure at the bottom of the DC potential well.  The 
    ejected bunched beam will have an energy spread of $5-10~\mathrm{eV}$ 
    and a time spread of $1.0-1.5~\mu\mathrm{s}$.\label{fig:new-RFQ}}
\end{figure}

In addition to production and other simulation work, there has been 
significant construction and testing on critical hardware for the 
\tamutrap{} experiment.  A prototype version of the RFQ cooler and buncher, 
including most ancillary systems and equipment, was assembled and optimized 
in continuous mode using an offline ion source.  The total transmission was 
found to be on the order of 25\% in this mode.  This relatively low efficiency 
is dominated by variations in the electrode distances; we are designing a new 
support and alignment structure for the RFQ, as shown in 
Fig.~\ref{fig:new-RFQ}.  Commissioning of this much more robust and sturdy 
RFQ is planned to be completed within one year.  The electronics for the 
prototype RFQ were found to work according to specifications, yielding up 
to $160~\mathrm{V}$ peak-to-peak at frequencies between $0.5-1.5~\mathrm{MHz}$. 

In addition to the RFQ, a beam diagnostic station based on a Faraday cup 
and micro-channel plate (MCP) detector has been installed.  Testing of 
these devices is currently ongoing, using the same offline ion source as 
described above.  An emittance measurement add-on based on the pepper-pot 
technique has been fabricated, and will be implemented in the future.

The immediate outlook for the \tamutrap{} facility involves completing the 
redesign of the RFQ and performing more realistic simulations of the decays 
of interest, in addition to continuing development of various beamline 
systems.  In particular, simulations are planned with GEANT4, and timing 
and control systems for the beamline are being developed around a National 
Instruments FPGA controlled by LabView.

\section{Atom Trapping with \trinat}
Magneto-optical traps (MOTs) provide an excellent source of radioactive atoms 
which can be easily polarized using atomic techniques.  The isotope currently 
being studied by the \trinat{} collaboration is the $\tfrac{3}{2}^+
\rightarrow\tfrac{3}{2}^+$ mixed Fermi/Gamow-Teller isobaric analogue decay 
of \tsups{37}K.\ \ In this case, the angular distribution of the decay as 
given in Refs.~\cite{JTWa,JTWb} is much richer given that the nucleus may 
be polarized and/or aligned:
\begin{align}
  & dW = dW_\circ\xi\Bigg[
  1+a\frac{\vec{p}_e\cdot\vec{p}_\nu}{E_eE_\nu} + 
  b\frac{m_e}{E_e}  + \frac{\vec{I}}{I}\cdot\left[
    A_\beta\frac{\vec{p}_e}{E_e} + B_\nu\frac{\vec{p}_\nu}{E_\nu} + 
    D\frac{\vec{p}_e\times\vec{p}_\nu}{E_eE_\nu} \right]\nonumber\\
  &\qquad\quad\quad + c_\mathrm{align}\bigg\{\left[
    \frac{\vec{p}_e\cdot\vec{p}_\nu}{3E_eE_\nu} - 
    \frac{(\vec{p}_e\cdot{\hat{i}})(\vec{p}_\nu\cdot{\hat{i}})}{E_eE_\nu}\right]
  \times\bigg[
    \frac{I(I+1)-3\langle(\vec{I}\cdot\hat{i})^2\rangle}{I(2I-1)}\bigg]
  \Bigg\}\Bigg]\label{eq:JTW-mixed}
\end{align}
where again $dW_0$ is given by Eq.~\eqref{eq:beta-decay}.  Being a mixed decay,
the correlation parameters in this case are all functions of $\rho\equiv 
C_AM_{GT}/C_VM_F$ (see Table~\ref{table:37K-params}), where $C_A$ ($C_V$) are the 
axial-vector (vector) semi-leptonic form factors and $M_{GT}$ ($M_F$) are the 
Gamow-Teller (Fermi) matrix elements of the decay.  

\begin{table}\centering
  \begin{tabular}{r@{$\ $}l@{\,}r@{$.$}l}
    \hline\hline
     $a_{\beta\nu}=$  & $\frac{1-\rho^2/3}{1+\rho^2}$ 
     & $=\ \ 0$&$6580(61)$\\[0.25em]
     $b=$  & \multicolumn{3}{l}{$\!\!\!\!0$ (sensitive to 
       scalars and tensors)}\\[0.25em]
     $A_\beta=$ & $\frac{-2\rho\ }{1+\rho^2}\left(\sqrt{3/5}-\rho/5\right)$
     & $=-0$&$5739(21)$\\[0.25em]
     $B_\nu=$ &  $\frac{-2\rho\ }{1+\rho^2}\left(\sqrt{3/5}+\rho/5\right)$
     & $=-0$&$7791(58)$\\[0.25em]
     $c_\mathrm{align}=$ & $\frac{4\rho^2/5}{1+\rho^2}$
     & $=\ \ 0$&$2053(36)$\\[0.2em]
     $D=$ & \multicolumn{3}{l}{$\!\!\!\!0$ (sensitive to imaginary 
       couplings)}\\[0.15em]
     \hline\hline
  \end{tabular}
  \caption{Standard model predictions of the correlation parameter values for 
    the decay of \tsups{37}K.\ \  The uncertainties quoted result from the 
    $\pm1.2\%$ precision to which the observed $ft$ value determines 
    $\rho$.  \label{table:37K-params}}
\end{table}

\begin{figure}\centering
  \includegraphics[height=0.333\textheight]{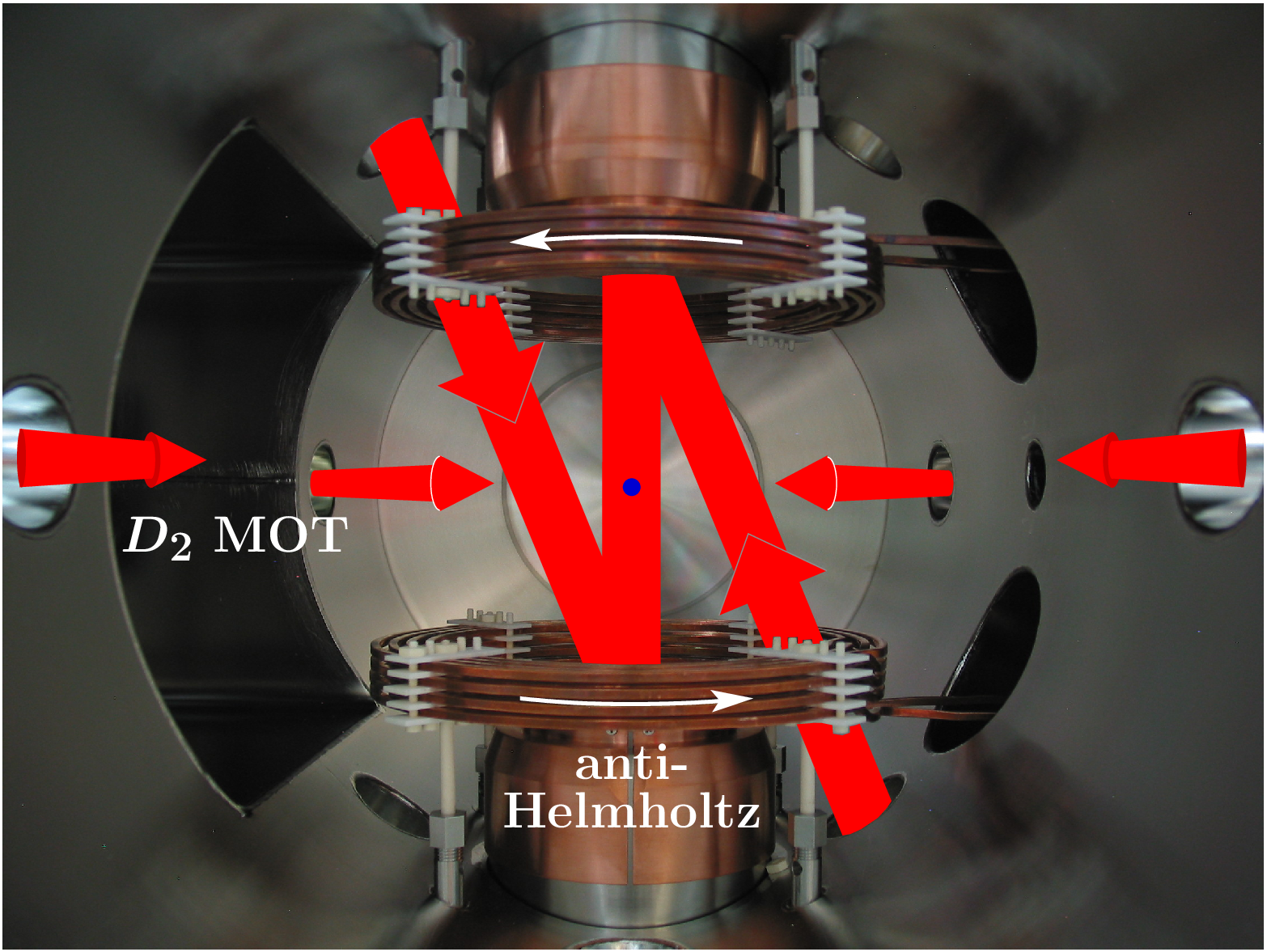}
  \ \ 
  \includegraphics[height=0.333\textheight]{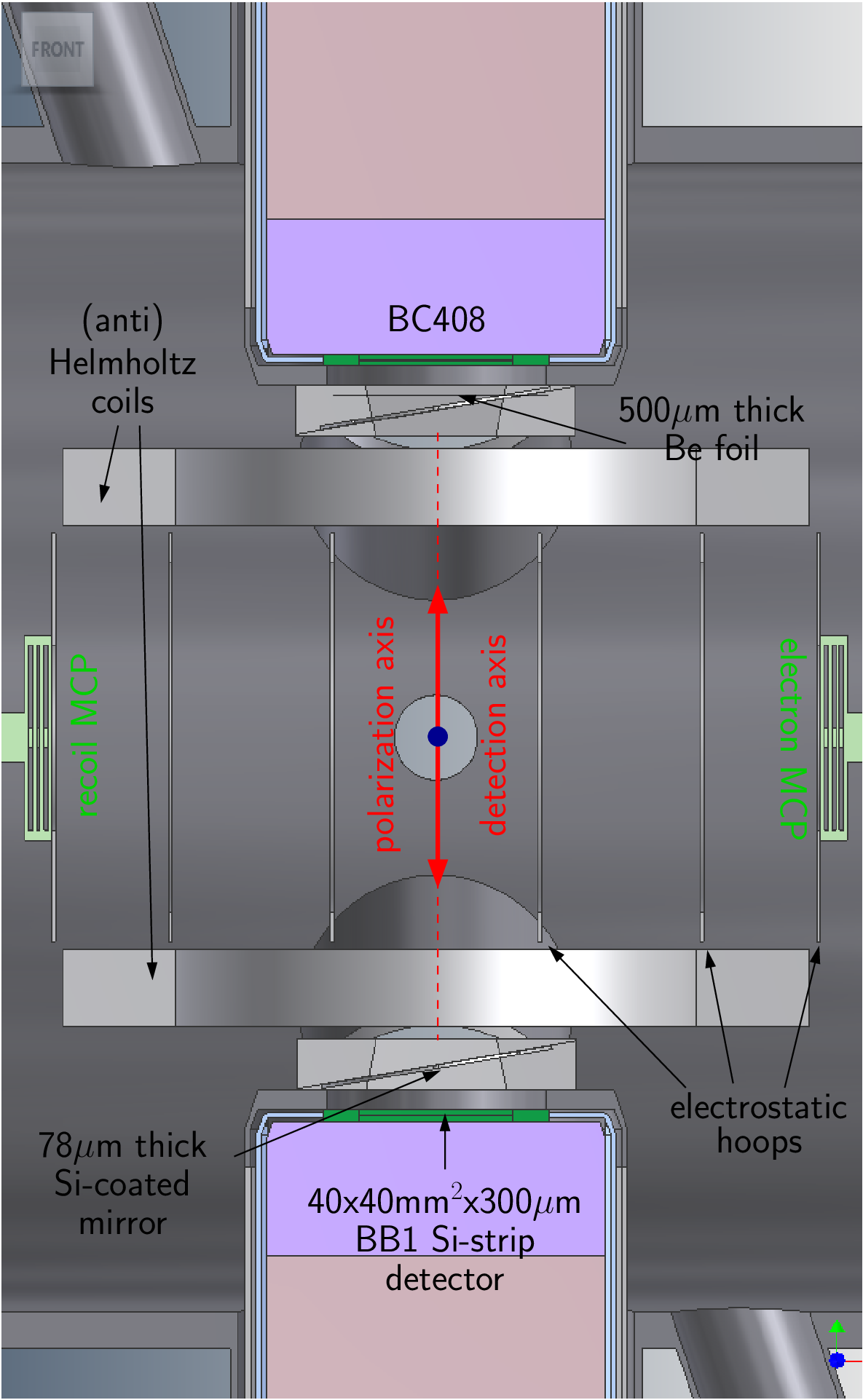}
  \caption{Side views of \trinat's detection chamber.  On the left is a picture 
    as viewed from the recoil MCP toward the electron MCP.  The two re-entrant 
    flanges house the $\beta$ telescopes.  Also visible are the water-cooled 
    leads for the coils which provide the Helmholtz (optical pumping) and 
    anti-Helmholtz (magneto-optical trapping) fields.  On the right is a 
    schematic diagram of a view rotated $-90^\circ$ from the picture on the 
    left.  See the text for a description of the elements shown.
    \label{fig:detection-MOT}}
\end{figure}

As in previous experiments, we continue to employ a double-MOT 
system~\cite{swan98} to increase efficiency and remove backgrounds from 
untrapped atoms.  Figure~\ref{fig:detection-MOT} shows the improved geometry 
we have recently commissioned which is optimized for measuring polarized 
correlation parameters.  
The two $\beta$-telescope detectors consist of a $300~\mu$m-thick 
double-sided Si-strip detector and a BC408 plastic scintillator; this will 
make the polarization axis -- the sign of which is determined by the 
polarization of the laser beams -- equivalent to the $\beta$ detection axis.  
This is an \emph{ideal} geometry for measuring the $\beta$ asymmetry, where 
one has the ability to flip the initial nuclear polarization on timescales 
of a few milliseconds.  The only material between the source and detectors 
is a thin mirror and a $0.1~\mathrm{mm}$ Be foil to separate the ultra-high 
vacuum of the chamber from the detector ports.  

On the right panel of Fig.~\ref{fig:detection-MOT} one can see the two 
micro-channel plate detectors we have placed; the view in the left pane 
is from one to the other.  Electrostatic hoops generate a uniform electric 
field whose function is threefold: (1) to increases collection efficiency 
of the recoiling \tsups{37}Ar\tsups{+} ions into the recoil MCP, (2) to help 
separate their different charge states of the recoiling ions, and (3) to 
accelerate the negatively charge shake-off electrons in the opposite direction 
onto an $e^-$ MCP.  The addition of a pulsed $355~\mathrm{nm}$ laser which 
is able to photoionize atoms in the $P_{1/2}$ excited state (but not the 
$S_{1/2}$ ground state) allows us to monitor the excited state population and 
cloud characteristics (size, position, temperature) in the following way: when 
a neutral atom is photoionized, an ion is produced which will be accelerated 
onto the position-sensitive recoil MCPs; gating on the time relative to the 
laser pulse results in a very clean, essentially background-free signal.

The MOT does not provide a polarized source of atoms, so we release the 
atoms, quickly polarize them using optical pumping techniques, make our 
asymmetry measurements, and then turn the trap back on to recollect the 
atoms before they expand too far and are lost.  Every $700~\mathrm{ms}$ 
we switch the polarization state of the optical pumping light and hence 
flip the nuclear spin of the \tsups{37}K.\ \ Unlike many other polarized 
experiments, we are able to measure the polarization of the cloud 
\emph{in situ} by fitting the observed $P_{1/2}$ excited state populations 
as a function of time to a optical-Bloch equation model of the optical pumping 
process.  
Using \tsups{41}K, a stable isotope which has a very similar electronic 
structure to \tsups{37}K, we have demonstrated the ability to polarize 
laser-cooled atoms to  $99.4\pm0.4\%$ nuclear polarization.  

We collected data for the first time with the new system in December 2012.  
Relative to the last precision measurement on radioactives, many extensive 
improvements were made to the system: new detection chamber, adopting an 
AC-MOT~\cite{harveyPRL2008} over the traditional DC-MOT to reduce eddy 
currents, installing new $\beta$ detectors, and upgrading from CAMAC to VME.
Given all these major changes, the goal of the December run was not only to 
show physics improvement by publishing a meaningful value for the $\beta$ 
asymmetry but to also to demonstrate engineering improvements and 
characterize the system that the collaboration will now use to produce more 
precise measurements in the future.  The whole system operated and ran in 
according to the design specification with the exception of the electrostatic 
field: sparking limited us to $350~\mathrm{V/cm}$ instead of our goal of 
$1~\mathrm{kV/cm}$.

\begin{figure}\centering
  \includegraphics[angle=90,width=0.6\textwidth]{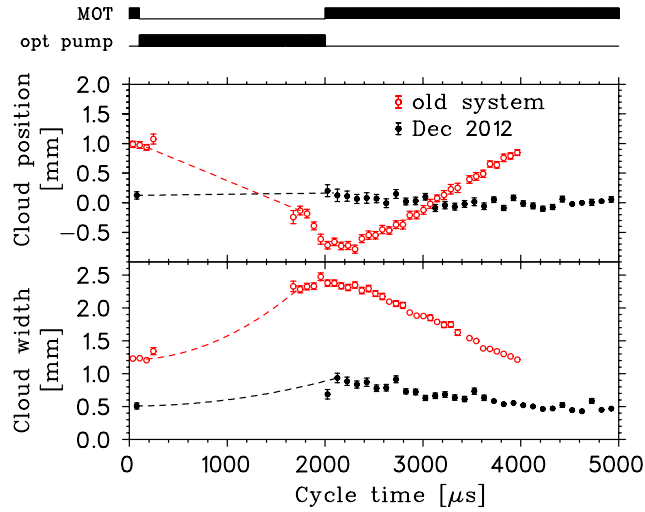}
  \caption{Position and size of the \tsups{37}K cloud as a function of time 
    in the trapping/polarizing cycle before and after improvements to the 
    system.  There are not enough photoions produced during optical pumping 
    times since so few atoms get excited to the $P_{1/2}$ state, whereas atoms 
    get excited often during MOT times.  One can clearly see that the 
    cloud position was much more stable in the latest run.  Furthermore, based 
    on the size and expansion of the cloud when the trapping MOT beams are off, 
    the cloud is half as large and there is a $4\times$ reduction in its 
    temperature.\label{fig:cloud-characteristics}}
\end{figure}

One clear indication of the improvements made is in the size and position of 
the cloud during our trapping/polarizing cycle.  As one can clearly see in 
Fig.~\ref{fig:cloud-characteristics} where we plot the cloud characteristics 
as determined by the photoion data, we have greatly improved our control and 
temperature of the atom cloud.  Plotted are the cloud position and size as 
determined by the position of photoions in the MCP detector.  We start with 
trapped atoms, release them by turning off the MOT beams and optically pump 
to polarize the laser-cooled cloud.  Before the cloud expands too much, we 
turn off the optical pumping beams and turn the MOT beams back on to re-trap 
atoms before they escape.  However, it is now clear that with the 1-inch 
diameter optical-pumping and MOT beams we have, we could have increased our 
duty cycle dramatically by reducing the time we spent re-trapping the atoms.

\begin{figure}\centering
  \includegraphics[angle=90,width=0.8\textwidth]{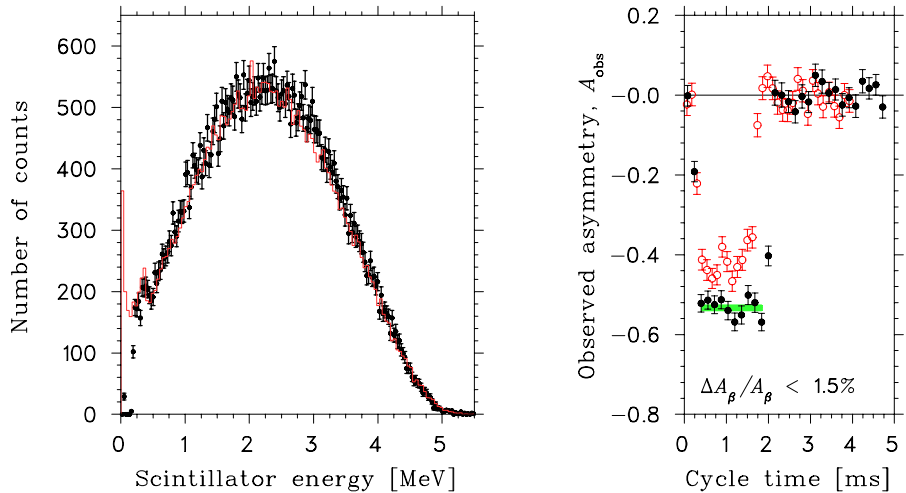}
  \caption{Energy spectrum in one of the scintillators (left) and preliminary 
    super-ratio asymmetry (right) for the Dec.\ 2012 run.  A {\scshape Geant}4 
    simulation of the $\beta$ spectrum is overlayed, with no background 
    subtracted from the data or added to the simulation, demonstrating we have 
    an extremely clean and well-understood system.  Analysis of 
    the super-ratio is still in progress, but preliminary analysis of the 
    Dec.\ 2012 data (filled circles) show a dramatically improved 
    asymmetry compared to the old system (open circles) when $B_\nu$ was 
    measured~\cite{melconianPLB}.\label{fig:geant-comparison}}
\end{figure}

Figure~\ref{fig:geant-comparison} shows the $\beta$ spectrum during polarized 
times ($300-2000~\mu\mathrm{s}$) in one of the scintillators requiring a 
coincidence with (\emph{i}) a $\beta$ in the $\Delta E$ Si-strip detector 
and (\emph{ii}) a shake-off electron in the $e^-$ MCP to ensure the decay 
occured from the trap (rather than, e.g., from the walls or the mirror in 
front of the telescope).  Requiring an anti-coincidence with the opposite 
$\beta$-telescope had a negligible effect on the spectrum.  The overlayed 
histogram is a {\scshape Geant4} simulation where no backgrounds or other 
effects have been added; the comparison is the raw data to the corresponding 
simulated data.  The analysis is preliminary at this stage, however one 
can already see that the spectrum--even the 511~keV Compton edge from the 
annihilation radiation--is reproduced \emph{extremely} well in the monte 
carlo.

The right panel of Fig.~\ref{fig:geant-comparison} shows a preliminary 
analysis of the asymmetry observed in the Dec.\ 2012 run via the 
``super-ratio'' technique.  The super-ratio, $R$, is defined in terms of the 
measured energy-dependent detector count rates for the two spin states, 
$r^\pm_i(E)$, to be
\begin{align}
  R=\frac{r_1^-(E_e)\,r_2^+(E_e)}{r_1^+(E_e)\,r_2^-(E_e)}.
\end{align}
Based on the angular distribution, Eq.~\eqref{eq:JTW-mixed}, the 
$\beta$ asymmetry parameter may be extracted from the super-ratio
according to
\begin{align}
  \frac{1-\sqrt{R}}{1+\sqrt{R}}=A_\mathrm{obs}(E_e)=\langle P\rangle 
  A_\beta\,\frac{v}{c}\langle\cos\theta\rangle,
\end{align}
where $P$ is the average nuclear polarization, $v/c$ is the velocity of the 
$\beta$, and $\langle\cos\theta\rangle$ is the average value of $\cos\theta$ 
integrated over the $\beta$-telescopes' angular acceptance.  This method for 
extracting $A_\beta$ is superior over a simple asymmetry because many 
systematic effects cancel to first order in the super-ratio, including 
spin-dependent and detector efficiencies. As expected, the observed asymmetry 
is large during polarized times and consistent with zero during trapping 
times.  The open circles represent the $\beta$ asymmetry observed in the old 
system when we measured the neutrino asymmetry~\cite{melconianPLB} before we 
added the shake-off electron detector; in that case the asymmetry was 
significantly attenuated from unpolarized atoms that did not decay from the 
trap, precluding us from making a precision measurement of $A_\beta$.  Our 
improved system ensures the $\beta$ decays occured from trap and already show 
a much higher asymmetry.  The statistical uncertainty in the observed 
asymmetry is below 1.5\%, however analysis of the systematics remains 
on-going.

\section{Conclusions}
Ion and atom traps provide a powerful tool for performing precision 
$\beta$-decay measurements of angular correlations.  These correlations are 
sensitive to the form of the weak interaction and may be used to complement 
searches for new physics at colliders if precisions of 0.1\% may be reached. 
We have presented progress on two efforts pushing the precision frontier as 
beyond the standard model tests:  the \tamutrap{} facility at the Cyclotron 
Institute which will measure $\beta-\nu$ correlations in proton-rich nuclei 
using a Penning trap; and the \trinat{} facility at \triumf{} which has 
recently measured the $\beta$ asymmetry in the mirror decay of \tsups{37}K.

\section{Acknowledgments}
We are grateful to the support staff of the Cyclotron Institute and of 
\triumf.  This work was supported by the U.S. Department of Energy's Grant 
No.\ DE-FG02-93ER40773 and Early Career Award ER41747.

\end{document}